# Sparse Matrix Multiplication on CAM Based Accelerator

Leonid Yavits and Ran Ginosar

**Abstract**— Sparse matrix multiplication is an important component of linear algebra computations. In this paper, an architecture based on Content Addressable Memory (CAM) and Resistive Content Addressable Memory (ReCAM) is proposed for accelerating sparse matrix by sparse vector and matrix multiplication in CSR format. Using functional simulation, we show that the proposed ReCAM-based accelerator exhibits two orders of magnitude higher power efficiency as compared to existing sparse matrix-vector multiplication implementations.

———————— ◆ ————————

## 1 INTRODUCTION

Sparse matrix multiplication is a frequent bottleneck in large scale linear algebra applications, especially in data mining and machine learning [16]. The efficiency of sparse matrix multiplication becomes even more relevant with the emergence of big data, giving rise to very large vector and matrix sizes.

A substantial body of literature explores sparse matrix multiplication optimization techniques. The prior work can be divided into three categories based on implementation platform (general purpose CPU or multicore, GPU and dedicated hardware accelerators), as summarized in TABLE 1. However the majority of previous studies target sparse matrix by dense vector multiplication (SpMV) or sparse matrix by dense matrix multiplication (SpMM). Sparse matrix by sparse vector or matrix multiplication (dubbed SpMSpV and SpMSpM in this paper) have rarely been addressed [18].

SpMV and SpMM optimization techniques use the column index of a multiplier (left) matrix element to explicitly address the multiplicand (right) vector (matrix) element in memory [13][16]. The uniqueness of SpMSpV and SpMSpM is that both multiplier matrix and multiplicand vector (matrix) are sparse and are frequently stored in Compressed Sparse Row (CSR) format. In CSR, using the column index of the (left) matrix element to directly access its counterpart is impossible. Instead, the (left) column index has to be matched to the (right) row index. In the CAM-based architecture, index matching is performed efficiently by the CAM; a successful match activates the correct row in the adjacent RAM, retrieves the corresponding nonzero (right) element, and the two elements are multiplied.

In this paper, we present the CAM-based accelerator architecture and evaluate its performance and power consumption. We find that our CAM based CSR SpMSpV accelerator outperforms GPU and multicore based implementations of SpMV and SpMM, while achieving much higher power efficiency. We show that a Resistive RAM/ CAM based implementation of the CSR SpMSpV accelerator leads to significant area saving.

TABLE 1
PRIOR WORK ON SPMM

| Category | Existing Work |
|---|---|
| General Purpose Processors | Off-the-shelf [21] |
| | Advanced multicore [22] |
| GPU | [8][14][16] |
| Dedicated Hardware Solutions | FPGA [13] |
| | Manycore Processor [15] |
| | Distributed Array Processor [7] |
| | Systolic Processor [17] |
| | Associative accelerator using STT-MRAM based TCAM [5] |
| | 3D LiM [18] |

The rest of this paper is organized as follows. Section 2 presents the architecture of the SpMSpV accelerator. Section 3 discusses the resistive implementation. Section 4 details the evaluation methodology and presents the simulation results. Section 5 offers conclusions.

## 2 CAM-BASED SPMSPV ACCELERATOR

In this section, we introduce the CSR SpMSpV accelerator, present the CSR SpMSpV algorithm, and explore the design parameters.

### 2.1 Proposed Architecture

The architecture of the CSR SpMSpV accelerator is presented in Fig. 1. The accelerator consists of $k$ identical *acceleration modules* each containing a Content Addressable Memory (CAM) array, juxtaposed with a RAM array, and a floating-point multiplier. The number of modules $k$ is a design parameter defined in Section 2.3. The inputs of each module come from memory. The outputs of the acceleration modules are connected to a floating-point accumulator ACC, which sums up the results of multiplications (the

———————————————

* *Leonid Yavits, E-mail: yavits@technion.ac.il.*
* *Ran Ginosar, E-mail: ran@ee.technion.ac.il.*
 *(\*)Authors are with the Department of Electrical Engineering, Technion-Israel Institute of Technology, Haifa 32000, Israel.*

singleton products). The output of the accumulator is fed to memory.

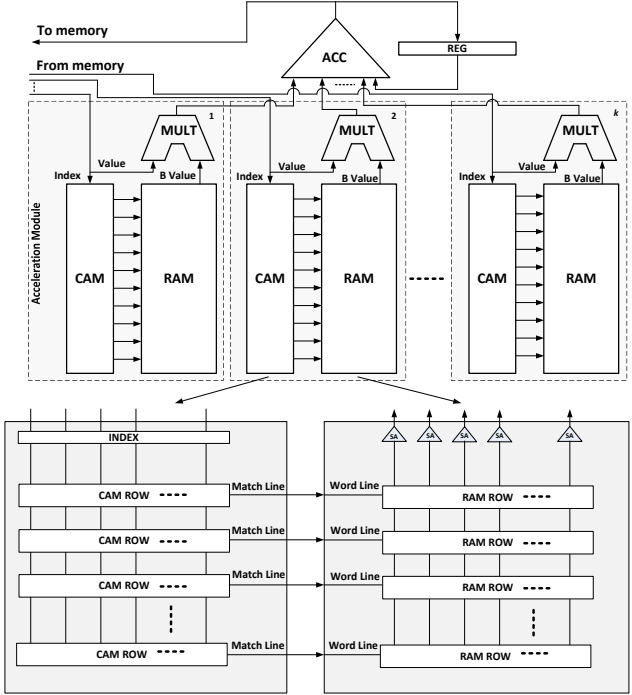

Fig. 1. The proposed SpMSpV Accelerator Architecture

The CAM consists of an INDEX register that holds the comparison data pattern (the column index) and an array of CAM rows (Fig. 1). On a match of the index and one of the rows, the corresponding match line is set. The match lines are fed into the juxtaposed RAM array as word lines. Thus, a match in the CAM selects one word in the RAM.

The CAM-RAM pair supports two operations: (1) initialization and (2) search-and-read. During initialization, the nonzero elements of the sparse (right) vector (designated *B*) are written (from memory) into the RAM, and their corresponding indices are written into the CAM, so that the nonzero element and its index are stored in corresponding rows.

During search-and-read, the column index of a nonzero element of the (left) sparse matrix (designated *A*) is placed in the INDEX register and *compare* is performed. The matching row selects the corresponding word in the RAM and the selected nonzero B element is read from the RAM into the multiplier.

The CSR SpMSpV accelerator matches $k \cdot h$ vector indices (where $h$ is the height of the CAM/RAM array), and multiplies and accumulates $k$ pairs of matrix/vector elements in a single clock cycle. Hence its peak performance is $k \cdot h$ index matching OPs and $2k$ FLOPs per cycle.

## 2.2 Algorithm

The CSR SpMSpV algorithm implemented by the accelerator is presented in Fig. 2. We assume that the sparse data is stored in Compressed Sparse Row (CSR) format [16].

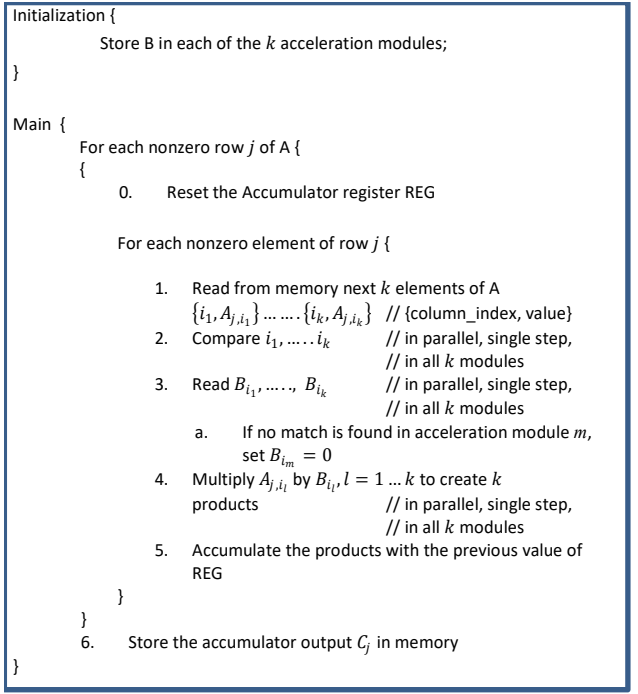

Fig. 2. The SpMSpV algorithm

The algorithm is first explained and then demonstrated by means of Fig. 3. The algorithm has two stages. The first stage is the initialization, where the sparse multiplicand vector B is stored in each acceleration module (so that there are $k$ copies of the vector B in the local memory of the accelerator). The nonzero values are stored in RAM and the corresponding column indices are stored in CAM, in the corresponding rows. The initialization step does not need to be repeated as long as different sparse multiplier matrices are multiplied by the same multiplicand vector.

The second stage is the main multiplication procedure. It is performed serially for all nonzero rows of the sparse matrix A (with the exception of diagonal matrices where the procedure can be performed in parallel for a number of nonzero rows to speed up the multiplication). The initial step (step 0) resets the accumulator register REG (*cf*. Fig. 1). The internal nested loop (steps 1 through 5) is repeated $\lceil nzr_j/k \rceil$ times, where $nzr_j$ is the number of nonzero elements in the $j^{th}$ row. In step 1, the next $k$ nonzero elements of the $j^{th}$ row of multiplier matrix A are read from memory. In step 2, the column index of each element is compared by the CAM with the entire set of indices of vector B. The matching row in each acceleration module is tagged, and the corresponding word of the juxtaposed RAM array is selected. In step 3, up to $k$ nonzero elements of vector B are read from the RAM arrays. If there is no match in a CAM array, meaning that there is no nonzero element of B matching the nonzero element of the $j^{th}$ row of A, the RAM array outputs a '0'. In step 4, the nonzero A elements are multiplied by the corresponding nonzero B elements, creating up to $k$ singleton products. In step 5, these singleton products are summed up by the accumulator ACC, together with the saved sum of the previous iteration.

After $\lceil nzr_j/k \rceil$ iterations of the internal loop, the $j^{th}$ element of the product vector C is ready. If $C_j \neq 0$, it is stored in memory, along with its index $j$.

The external loop is repeated until all nonzero rows of the multiplier matrix A are processed. Every step of the SpMSpV algorithm takes a single clock cycle. All operations are pipelined, so that the SpMSpV cycle time is $O(nnz/k)$ where $nnz$ is the number of nonzero elements in matrix A.

One iteration of SpMSpV is exemplified in Fig. 3. Four nonzero elements of a row of A and their column indices are fetched, (4,56), (10,16), (12,78) and (20,12). The acceleration modules identify that 98, 40 and 32 exist in rows 4, 10 and 12 of B, respectively, and that row 20 has no matching nonzero element. The three values are extracted and four singleton products are computed, 56×98, 16×40, 78×32 and 0, followed by their accumulation.

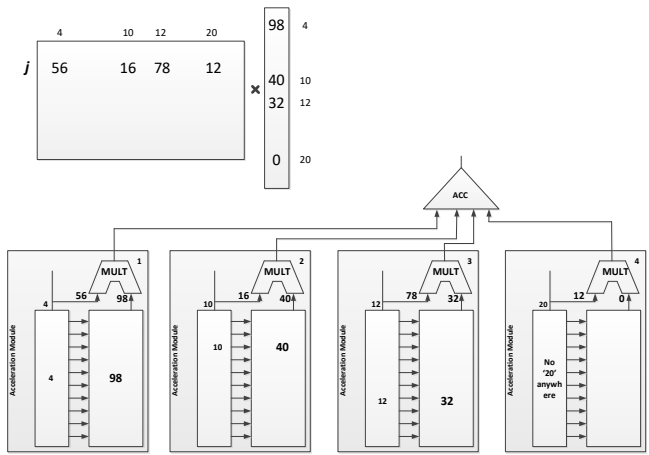

Fig. 3. SpMSpM example

The SpMSpM can be performed on the SpMSpV accelerator column by column, with a column of the sparse matrix B replacing the sparse vector B.

Although the accelerator is optimized for SpMSpV, it can also implement dense by dense and sparse by dense matrix and vector multiplication as well. However, while the same performance can be achieved, the area and power figures would be suboptimal.

### 2.3 Design Space Exploration

The performance of the accelerator is defined by the number of acceleration modules $k$, and by memory bandwidth. Assume that the sparse vector B is pre-stored in the RAM arrays of the acceleration modules. Up to $nzr_j$ multiply-and-accumulate operations can be performed per $nzr_j + 1$ memory accesses (reading $nzr_j$ nonzero elements of a row of A and writing the result). The number of acceleration modules $k$ should not exceed the number of nonzero elements (and their indices) that can be fetched in a cycle. SpMSpM may reach a higher peak performance of $L \times nzr_j$ multiply-accumulate operations per $nzr_j + 1$ memory accesses (where $L$ is the horizontal dimension of the multiplicand matrix).

The height $h$ of the CAM/RAM arrays is defined by the number of nonzero elements in sparse vector B. However, the multiplication by a larger vector is possible, by iterating the algorithm of Fig. 2 over a $h$-size interval of B until the entire vector is processed, and updating the elements of the product vector C accordingly. Hence the maximum possible number of nonzero elements in sparse vector B does not need to be fixed at design time. The maximum height of the CAM/RAM arrays $h$ is rather derived from the area and power budget of the accelerator.

The width of the CAM array $w$ is defined at design time as $w = \log_2 N$, where $N$ is the max length of sparse vector B.

The data item width is defined by the wordlength of the nonzero value (32 bits, assuming floating point) and column index ($w$ bits).

Fig. 4 shows the number of acceleration modules (a) and peak performance of the SpMSpV accelerator (b) as a function of memory bandwidth, assuming operating frequency of 2 GHz, $h = 2^{20}$ and $w = 32$. Memory bandwidth of the state of art high performance processors reaches above 250 GByte/sec [20], enabling 15 acceleration modules and peak SpMSpV integer index matching and floating-point performance of 30 PetaOP/s and 60 GFLOP/s respectively. However, emerging technologies such as 2.5D and 3D DRAM integration may enable higher memory bandwidth and consequently a larger number of acceleration modules and higher accelerator performance.

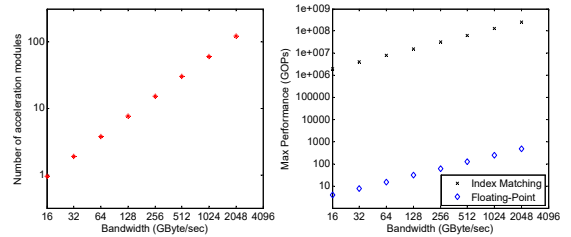

Fig. 4. Sensitivity to memory bandwidth: (a) Number of acceleration modules, (b) Peak performance (Integed index matching and Floating point)

## 3 RESISTIVE IMPLEMENTATION

Using the silicon area figures for CAM [10], RAM and floating point unit [1], and assuming the number of acceleration modules $k = 15$ and CAM/RAM array height $h = 2^{20}$, we estimate the area of the CMOS SpMSpV accelerator at $90 mm^2$ in 22nm technology node.

As CMOS feature scaling slows down, conventional memory technology experiences scalability problems. In response, resistive memory technologies are explored, *e.g.* Resistive RAM (ReRAM) based CAM [4][6], that can serve as scalable, long-term alternatives to CMOS CAM. Resistive memories store information by modulating the resistance of nanoscale storage elements (sometimes called memristors), and are expected to scale to smaller geometries. Resistive memories are non-volatile, which provides near-zero leakage power. However, ReRAM suffers from finite endurance, as compared to CMOS memories.

Memristors are two-terminal devices, where the resistance of the device is changed by the electrical current or voltage. The resistance of the memristor is bounded by a minimum resistance $R_{ON}$ (low resistive state, considered for digital memories as logic '1') and a maximum resistance $R_{OFF}$ (high resistive state, logic '0').

In this work, we introduce a resistive CAM/RAM (ReCAM/ReRAM) array (Fig 5(a)), where each ReCAM bitcell (Fig 5(b)) consists of a pair of ReRAM bitcells (Fig 5(c)), formed by a nonlinear bipolar memristor that effectively has a diode for preventing sneak paths [4]. The second memory bit of the ReCAM cell serves as a complementary bit.

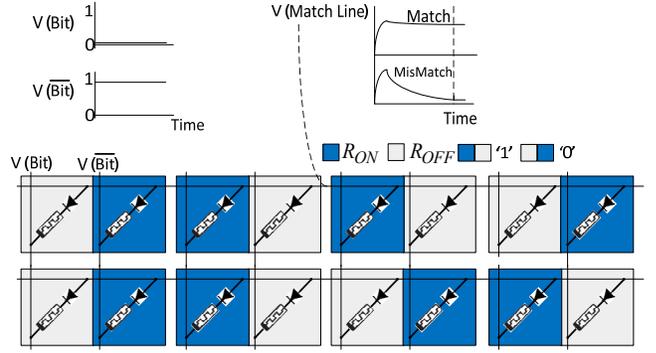

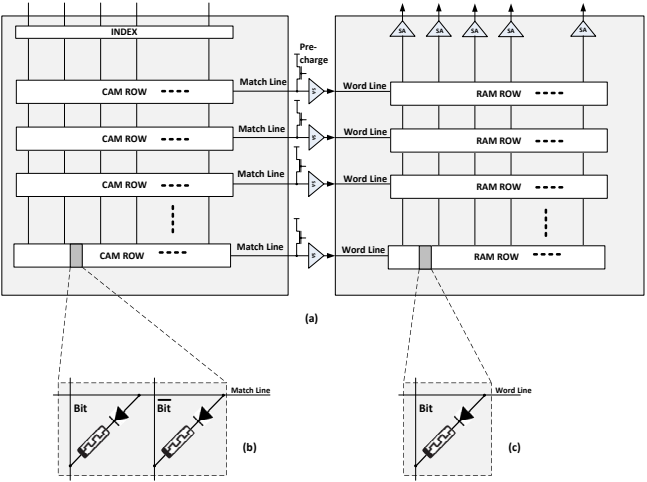

Fig 5. (a) ReCAM/ReRAM array, (b) ReCAM bitcell, (c) ReRAM bitcell

Compare operation in ReCAM is similar to compare in CMOS CAM [12]. The Match line is precharged and the key is set on Bit and Bit-not lines. In the columns that are ignored during comparison, the Bit and Bit-not lines are kept floating. If all unmasked bits in a row match the key (*i.e.*, when Bit line '1' is applied to an $R_{ON}$ memristor and Bit-not line '0' is applied to an $R_{OFF}$ memristor, or vice versa), the Match line remains high and '1' is supplied to the juxtaposed Word line of the ReRAM. If at least one bit is mismatched, the Match line discharges through an $R_{OFF}$ memristor and '0' is supplied to the corresponding ReRAM Word line.

Compare is illustrated in Fig 6, which shows a fragment of ReCAM storing '0110' in the first row and '0101' in the second row; The ReCAM content is compared with the '0110' index.

In ReCAM, sneak currents affect the compare operation (rather than read operation in a typical ReRAM crossbar). More specifically, there are sneak paths leading from a matching Match/Word Line (which is supposed to retain '1') through neighboring mismatching Match/Word Lines to the ground. The purpose of per-cell diode [4] is to terminate such sneak path, so that current can only flow from a Match/Word Line to the ground (through a Bit Line) in one direction.

ReRAM and ReCAM performance and energy figures are obtained by SPICE simulations [12]. Those energy figures are used in functional simulation of the SpMSpV simulator described in Section 4.

Fig 6. Compare in ReCAM

Memristor design enables ReCAM and ReRAM bitcells of $8F^2/l$ (where $l$ is the number of vertically integrated memristor layers [4]) and $4F^2$ footprint respectively. Such level of integration allows reducing the SpMSpV accelerator size by almost 30x to around $3mm^2$ in 22nm technology.

The switching time of memristor is in the range of a hundred picoseconds [2] allowing GHz SpMSpV accelerator operation. The energy consumption during compare is less than 1fJ per bit. Another factor to be aware of is the endurance of resistive memory which is in the range of $10^{12} - 10^{15}$ [9]. However since write into resistive memory is relatively scarce in the proposed architecture (the memory is only written when the vector (matrix) B needs to be updated), endurance is not a limiting factor.

## 4 EVALUATION

To simulate the SpMSpV accelerator, we use 640 square matrices with the number of nonzero elements spanning from hundred thousand to eight million, randomly selected from the collection of sparse matrices from the University of Florida [23]. A row of each of the 640 matrices, extracted from it in a random manner, is used as a multiplicand sparse vector. The maximum number of nonzero elements in such vector is 390. Consequently, the height of the simulated CAM/RAM array is set at $h = 512$.

Fig. 7 presents (a) the integer (index matching) and floating point performance, and (b) power efficiency of the SpMSpV accelerator simulated for $k = 15$ modules. The SpMV performance and power efficiency of NVidia's K20 [3][20] and SpMV performance of Intel's Xeon Phi [3] are also shown for comparison. The floating-point performance of SpMSpV is limited by the peak performance of 60 GFLOP/s, as defined in Section 2.3. The spread in both performance and power consumption of SpMSpV occurs because the number of nonzero elements per row is rarely a multiple of $k$. The SpMSpV accelerator outperforms state of art GPU and multicore, as well as the Associative Processor based [11] SpMV implementations. Since the size of CAM/RAM array is quite limited ($h = 512$), total power of SpMSpV accelerator is dominated by floating point operations and does not exceed 0.3W.

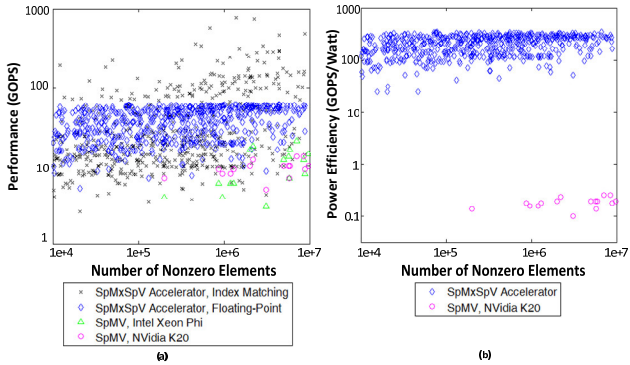

Fig. 7. Resistive SpMSpV accelerator (a) performance, (b) power efficiency; $k = 15$, $h = 512$

The SpMV power efficiency of advanced contemporary GPUs such as NVidia's K20 and GTX660 is in the 0.1-0.5 GFLOPs/W range [19][20]. A wide variety of multicore processors reportedly reach the SpMV power efficiency of up to 0.03 GFLOPs/W [22]. SpMSpV is more power hungry than SpMV since it requires massive index matching. Still, the power efficiency of the SpMSpV accelerator is superior to state of art GPU and multicore solutions.

## 5 CONCLUSIONS

Sparse matrix multiplication is of great importance for many linear algebra applications, especially machine learning. The efficient implementation of sparse matrix multiplication becomes even more critical when applied to big data problems.

The uniqueness of CSR SpMSpV and SpMSpM is in that unlike SpMV and SpMM, the column index of a multiplier (left) matrix element cannot be used to address the multiplicand (right) vector element. To pair the elements of the sparse matrix and sparse vector (matrix) stored in CSR format, their column and row indices have to be matched. We solve this issue by using juxtaposed CAM and RAM arrays, where CAM is used to match the indices, and select the RAM row where the corresponding nonzero element of the sparse multiplicand vector (matrix) is stored.

In this work, we explore the Resistive CAM/RAM based SpMSpV accelerator and evaluate its floating-point and index matching performance and power consumption. We find that such accelerator outperforms conventional GPU and multicore based implementations of SpMV and SpMM, while providing considerably better power efficiency. Additionally, we show that the resistive implementation leads to significant area saving.

## ACKNOWLEDGMENT

This research was partially funded by the Intel Collaborative Research Institute for Computational Intelligence and by Hasso-Plattner-Institut.

## REFERENCES

[1] A. Pedram, "Algorithm/architecture codesign of low power and high performance linear algebra compute fabrics", PhD Dissertation, the University of Texas at Austin, 2013
[2] A. Torrezan et al. "Sub-nanosecond switching of a tantalum oxide memristor." Nanotechnology 22.48 (2011): 485203.
[3] E. Saule, et al. "Performance Evaluation of Sparse Matrix Multiplication Kernels on Intel Xeon Phi." arXiv preprint arXiv:1302.1078 (2013).
[4] F. Alibart, T. Sherwood, D. Strukov. "Hybrid CMOS/nanodevice circuits for high throughput pattern matching applications", IEEE Conference on Adaptive Hardware and Systems, 2011
[5] G. Qing, X. Guo, R. Patel, E. Ipek, E. Friedman. "AP-DIMM: Associative Computing with STT-MRAM," ISCA 2013.
[6] J Li, et al. "1Mb 0.41 µm 2 2T-2R cell nonvolatile TCAM with two-bit encoding and clocked self-referenced sensing", IEEE Symposium on VLSI Circuits, 2013.
[7] J. Andersen, G. Mitra, D. Parkinson. "The scheduling of sparse matrix-vector multiplication on a massively parallel DAP computer." Parallel Computing 18, no. 6 (1992): 675-697.
[8] J. Bolz, I. Farmer, E. Grinspun, and Peter Schröoder. "Sparse matrix solvers on the GPU: conjugate gradients and multigrid." In ACM Transactions on Graphics, vol. 22, no. 3, pp. 917-924. ACM, 2003.
[9] K. Eshraghian, et al. "Memristor MOS content addressable memory (MCAM): Hybrid architecture for future high performance search engines", IEEE Transactions on VLSI Systems, 19.8 (2011): 1407-1417.
[10] L. Yavits, A. Morad, R. Ginosar, "Computer Architecture with Associative Processor Replacing Last Level Cache and SIMD Accelerator", IEEE Transactions on Computers, 2013
[11] L. Yavits, A. Morad, R. Ginosar, "Sparse Matrix Multiplication on Associative Processor", IEEE Transactions on Parallel and Distributed Systems, 2014
[12] L. Yavits, et al. "Resistive Associative Processor", IEEE Computer Architecture Letters, 2014
[13] L. Zhuo, V. Prasanna. "Sparse matrix-vector multiplication on FPGAs." International symposium on FPGA, ACM/SIGDA, 2005.
[14] M. Baskaran, R. Bordawekar. "Optimizing sparse matrix-vector multiplication on GPUs using compile-time and run-time strategies." IBM Research Report, RC24704 (W0812-047), 2008.
[15] M. Misra et al. "Efficient VLSI implementation of iterative solutions to sparse linear systems." Parallel Computing 19, no. 5 (1993): 525-544.
[16] N. Bell, M. Garland. "Implementing sparse matrix-vector multiplication on throughput-oriented processors." Conference on High Performance Computing Networking, Storage and Analysis, p. 18. ACM, 2009.
[17] O. Wing, "A content-addressable systolic array for sparse matrix computation." Journal of Parallel and Distributed Computing 2, no. 2, 1985.
[18] Q. Zhu, et al. "Accelerating sparse matrix-matrix multiplication with 3d-stacked logic-in-memory hardware", IEEE High Performance Extreme Computing Conference, 2013.
[19] R. Dorrance et al., "A scalable sparse matrix-vector multiplication kernel for energy-efficient sparse-BLAS on FPGAs", Proceedings of the 2014 ACM/SIGDA international symposium on Field-programmable gate arrays (pp. 161-170).
[20] S. Keckler, et al. "GPUs and the future of parallel computing", IEEE Micro, 31.5 (2011): 7-17
[21] S. Toledo, "Improving the memory-system performance of sparse-matrix vector multiplication." IBM J. Res. Dev. 41, no. 6 (1997): 711-725.
[22] S. Williams et al., "Optimization of sparse matrix–vector multiplication on emerging multicore platforms", Parallel Computing Journal, 35, no. 3 (2009): 178-194.
[23] T. Davis, Y. Hu, "The University of Florida sparse matrix collection," ACM Transactions on Mathematical Software (TOMS), 38, no. 1 (2011): 1.